\documentclass[aps,twocolumn,prl,showpacs,amsmath,amssymb,10pt]{revtex4-1}
\usepackage{graphicx}
\usepackage{scrextend}
\usepackage{manfnt,pifont}

\newcommand\figr[1]{Fig.\,\ref{#1}}

\newcommand\noi{\noindent}
\newcommand\AdS{{\rm AdS}}
\newcommand\Tr{{\rm Tr}}
\newcommand\inv{^{-1}}
\newcommand\zb{\bar{z}}
\newcommand\del{\partial}

\newcommand\cG{{\cal G}}

\newcommand\cN{{\cal N}}
\newcommand\cO{{\cal O}}

\newcommand{\be}{\begin{equation}}
\newcommand{\ee}{\end{equation}}
\newcommand{\bea}{\begin{eqnarray}}
\newcommand{\eea}{\end{eqnarray}}
\newcommand\restr[2]{{\left.\kern-\nulldelimiterspace#1\vphantom{\big|}\right|_{#2}}}

\newcommand\eg{\emph{e.g.}}

\begin{document}
\title{The ${\mathcal N}=4$ Superconformal Bootstrap}
\author{Christopher Beem,$^\text{\ding{113}}$ Leonardo Rastelli,$^\text{\mancube}$ and Balt C. van Rees$^\text{\manimpossiblecube}$} 
\affiliation{
\ding{113} Simons Center for Geometry and Physics, SUNY, Stony Brook, NY 11794-3636, USA\\
\mancube\,\,C.~N.~Yang Institute for Theoretical Physics, SUNY, Stony Brook, NY 11794-3840, USA
}

\begin{abstract}
\noi We implement the conformal bootstrap for $\cN=4$ superconformal field theories in four dimensions. Consistency of the four-point function of the stress-energy tensor multiplet imposes significant upper bounds for the scaling dimensions of unprotected local operators as functions of the central charge of the theory. At the threshold of exclusion, a particular operator spectrum appears to be singled out by the bootstrap constraints. We conjecture that this extremal spectrum is that of $\cN=4$ supersymmetric Yang-Mills theory at an S-duality invariant value of the complexified gauge coupling.

\end{abstract}
\pacs{11.25.Hf, 11.30.Pb;\qquad Preprint numbers: YITP-SB-13-10}
\maketitle

{\bf Introduction.}
It has become apparent in recent years that the conformal bootstrap program -- the analysis of conformally invariant quantum field theories using only consistency conditions from symmetries, unitarity, and associativity of the operator product expansion (OPE) -- can be implemented to great effect in any spacetime dimension to place substantial constraints on the spectrum of local operators \cite{Rattazzi:2008pe}. The constraints arising from a single four-point function are surprisingly powerful, leading to numerical bounds on scaling dimensions that appear to be saturated in known conformal field theories such as the two-dimensional minimal models \cite{Rychkov:2009ij} and the three-dimensional critical Ising model \cite{ElShowk:2012ht}. For a theory that lies at an exclusion threshold, it is in principle possible to recover the dimensions and three point functions of the operators appearing in the OPE decomposition of the four-point function \cite{Poland:2010wg,ElShowk:2012hu}. One expects these methods to be most effective in theories that are uniquely specified by a few basic properties, such as global symmetries. Maximally supersymmetric conformal field theories (SCFTs) are thus ideal candidates for such an analysis. 

In this letter we report initial results from the application of bootstrap methods to four-dimensional theories with $\cN=4$ superconformal invariance \footnote{A more complete account of our methods, along with additional results, will appear in \cite{BRR:long}.}. While $\cN=4$ supersymmetric Yang-Mills theory (SYM) has been the subject of intense study, all findings to date have either relied on perturbation theory or the planar limit, or they have used special simplifications that pertain only to BPS observables. We describe here the first nonperturbative results for the dimensions of unprotected operators at finite values of the central charge. We eschew any reference to a Lagrangian description of the theory and consider the universal four-point function of energy-momentum tensor multiplets, whose structure is strongly constrained by symmetry. We obtain upper bounds on the dimensions of unprotected operators of leading twist appearing in its OPE decomposition, and find strong indications that $\cN=4$ SYM does indeed saturate the bounds at a special point of its conformal manifold.

\begin{figure*}[ht!]
\includegraphics[width=167pt]{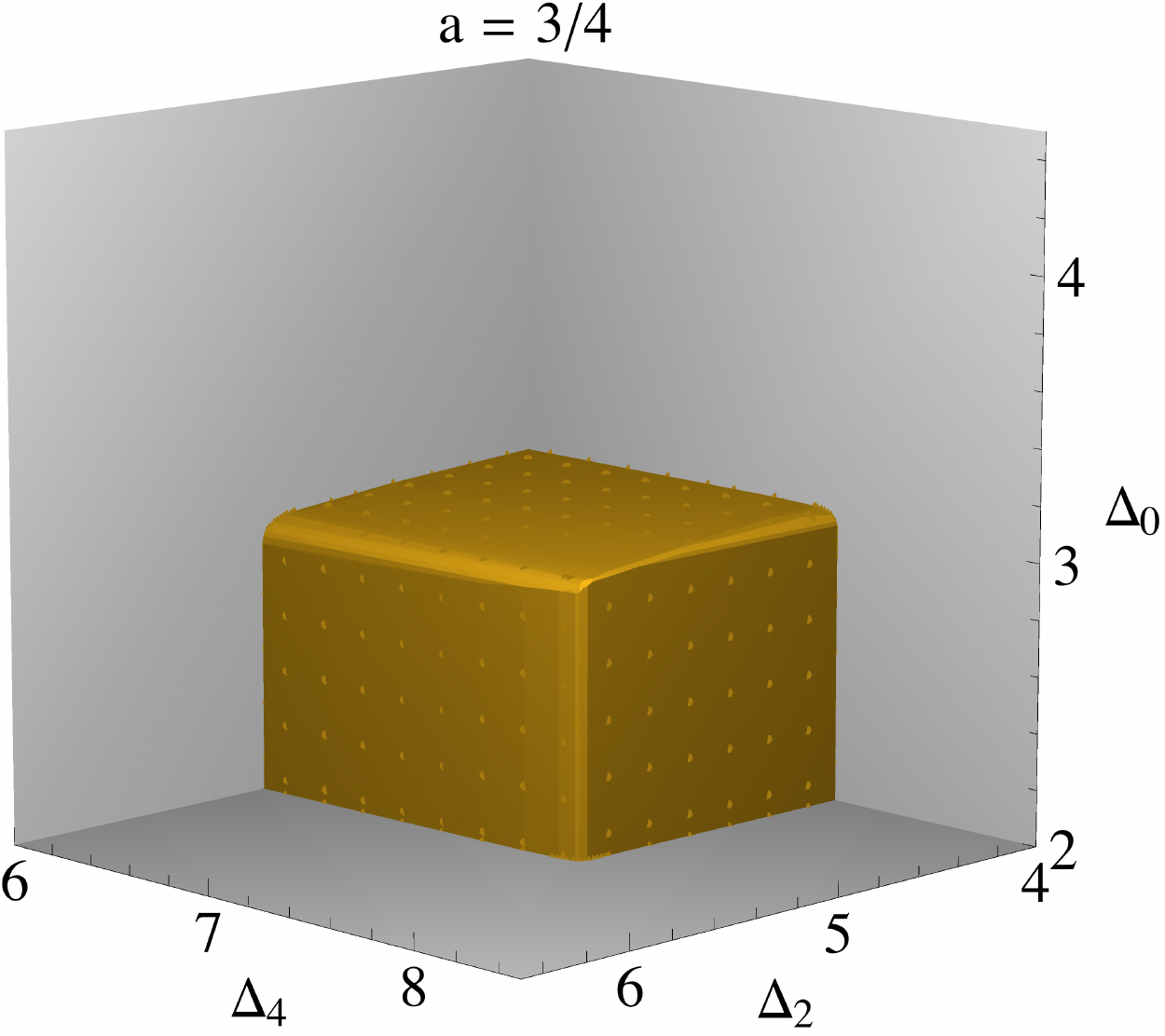}~
\includegraphics[width=167pt]{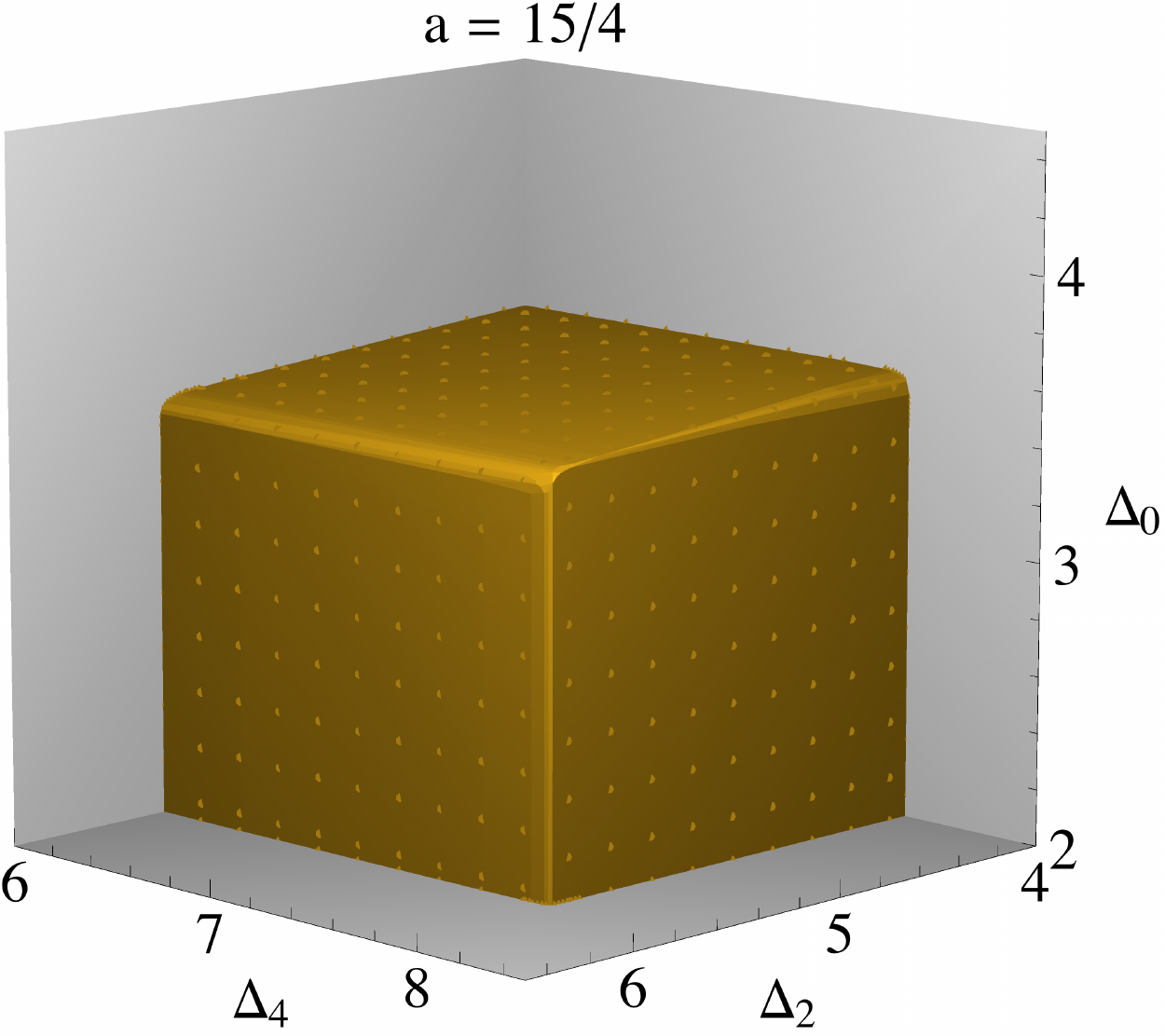}~
\includegraphics[width=167pt]{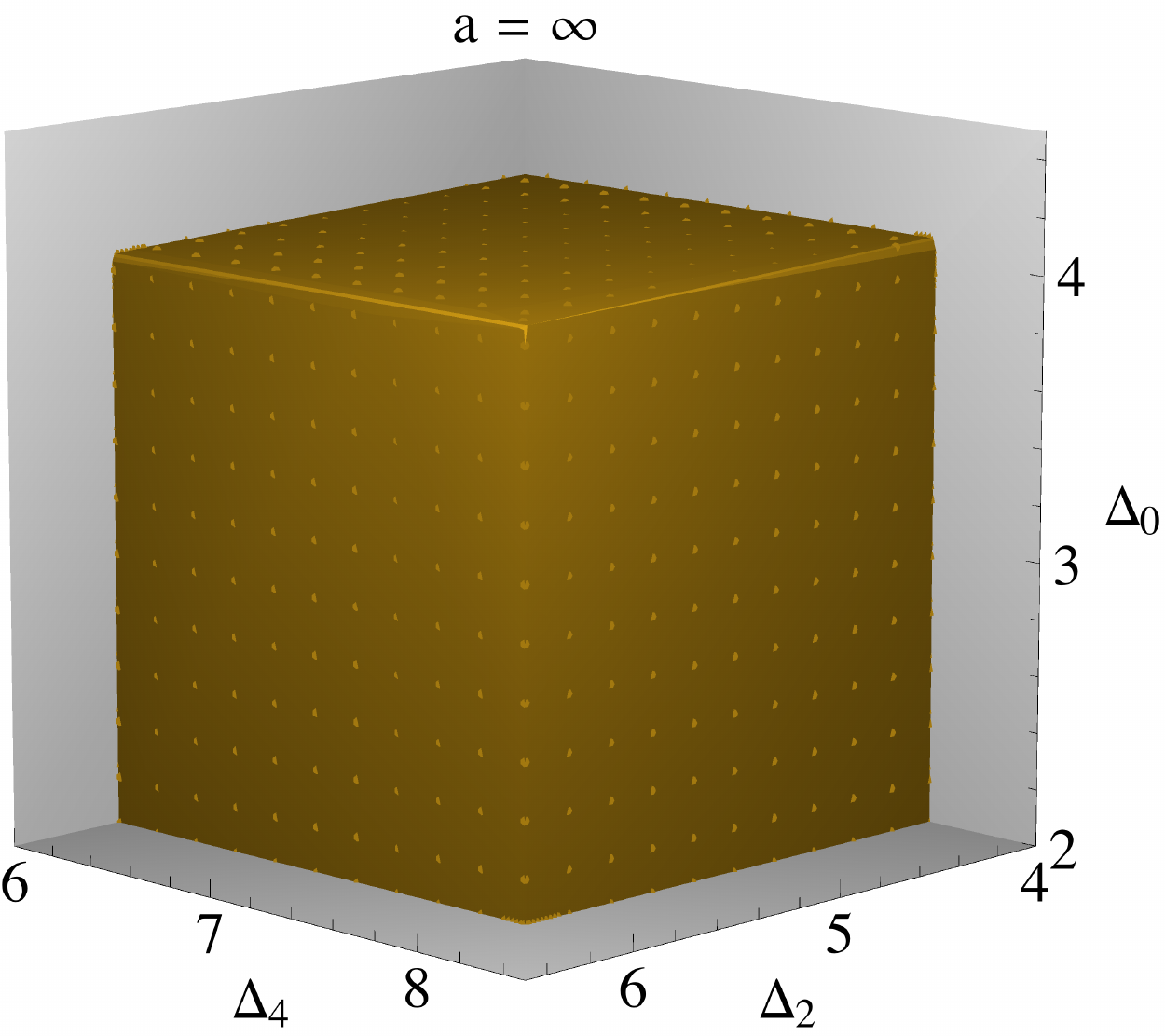}
\caption{Exclusion plots in the space of leading twist gaps $\Delta_0$, $\Delta_2$, and $\Delta_4$. Central charges $a=3/4$, $a=15/4$, and $a=\infty$ are shown, corresponding to $\cN=4$ SYM with gauge group $SU(2)$, $SU(4)$, and $SU(\infty)$, respectively. The area outside of a cube-shaped region is excluded.}
\label{fig:cubes} 
\end{figure*}

\vspace{10pt}
{\bf The sum rule.} 
In a four-dimensional $\cN=4$ SCFT, the energy-momentum tensor lies in a half-BPS multiplet whose superconformal primary, denoted here as  $\cO^{I}_{\bf 20^\prime}$, $I=1,\ldots,20$, is a scalar operator of dimension two that transforms in the $\bf 20^\prime$ representation of the $SU(4)_R$ R-symmetry group. Conformal symmetry fixes the form of its four-point function to
\be\label{4ptdef}\notag
\langle\,\cO_{\bf 20'}^{I_1}(x_1)\cO_{\bf 20'}^{I_2}(x_2)\cO_{\bf 20'}^{I_3}(x_3)\cO_{\bf 20'}^{I_4}(x_4)\,\rangle=\frac{A^{I_1I_2I_3I_4}(u,v)}{x_{12}^4x_{34}^4}~,
\ee
where we have introduced standard conformal cross-ratios $u=(x_{12}^2x_{34}^2)/(x_{13}^2x_{24}^2)$ and $v=(x_{14}^2x_{23}^2)/(x_{13}^2x_{24}^2)$. We will also make use of an alternate parameterization, $u= z\bar z,~v=(1-z)(1-\bar z)$ \cite{Freedman:1991tk,*Usyukina:1992jd,*Usyukina:1993ch,Dolan:2000ut}.

The constraints of superconformal invariance on this four-point function have been analyzed in detail in, \eg, \cite{Eden:2000bk,Eden:2001ec,Dolan:2001tt,Heslop:2002hp,*Dolan:2004mu,*Nirschl:2004pa}. We briefly review the results of \cite{Eden:2000bk,Dolan:2001tt} and proceed to derive the corresponding crossing symmetry sum rule. The function $A^{I_1I_2I_3I_4}(u,v)$ can be decomposed into channels corresponding to each representation of $SU(4)_R$ appearing in the tensor product $\bf 20^\prime\otimes\bf 20^\prime$, yielding six functions $A_R(u,v)$ for $R=\bf 1,\bf 15, \bf 20^\prime,\bf84,\bf105,\bf175$. Superconformal symmetry leads to relations among them, ultimately allowing the entire four-point function to be expressed algebraically in terms of a more elementary set of constituent functions. There is a single \emph{unprotected} function $\cG(u,v)$ that depends on the full spectrum of operators appearing in the OPE, along with two meromorphic functions and a constant that are \emph{protected}, and depend only on operators in the OPE that belong to shortened representations.

Crossing symmetry amounts to the invariance of the four-point function under the exchange $(x_1,I_1)\leftrightarrow(x_3,I_3)$, which implies several distinct relations among the $A_R(u,v)$. These relations can be processed in two steps. First, one extracts a closed set of equations for the protected functions only. Remarkably, those equations admit a unique family of solutions parametrized by the central charge appearing in the OPE of two stress tensors. In other words, the protected part of the amplitude is uniquely fixed by symmetry \footnote{The constraints of crossing symmetry on the spectrum of protected operators in ${\cal N}=4$ and ${\cal N}=2$ SCFTs is a beautiful subject that will be elaborated upon in \cite{miniboot}.}. Second, substituting the solution for the protected functions, there remains a single relation for $\cG(u,v)$ that is necessary and sufficient for crossing symmetry of the full four-point function,
\be\label{Gcross}\notag
v^2\cG(u,v)-u^2\cG(v,u)
+ 4(u^2-v^2)+\frac{4(u-v)}{a}=0~.
\ee
Here $a$ is the central charge, which for $\cN=4$ SYM with gauge group $G$ is given by $\dim G/4$.

To proceed, we consider the OPE decomposition of the crossing symmetry relation. This decomposition is relatively simple due to the relation $u^2\cG(u,v)= A_{\bf 105}(u,v)$ \cite{Arutyunov:2001mh}, which allows $\cG(u,v)$ to be expanded in conformal blocks for quasiprimary operators $\cO_{\bf 105}$ with nonzero three-point coupling $\langle\cO_{\bf 105}\cO_{\bf 20^\prime}\cO_{\bf 20^\prime}\rangle$. Superconformal symmetry restricts the possible multiplets in which such an operator can appear \cite{Eden:2001ec,Dolan:2001tt}: there are a few infinite families of multiplets obeying certain (semi-)shortening conditions, as well as long multiplets for which the superconformal primary is an $SU(4)_R$ singlet. The function $\cG(u,v)$ can be decomposed accordingly,
\be\label{Gparts}\notag
\cG(u,v)=\cG^{\rm short}(u,v)+\cG^{\rm long}(u,v)~  ,
\ee
and the two terms can be independently expanded,
\be\begin{split}\label{Gshortlong}\notag
&\cG^{\rm short} = \sum_{\ell} \left(g_{4,\ell}\,G^{(\ell)}_{\ell+4}(u,v) + g_{6,\ell}\,u G^{(\ell)}_{\ell+6}(u,v)\right)~,\\
&\cG^{\rm long} = \sum_{\ell=0,2,\ldots}\sum_{\Delta\geq2+\ell}\left(a_{\Delta,\ell}u^{\frac{\Delta-\ell}{2}}G^{(\ell)}_{\Delta+4}(u,v)\right)~.
\end{split}\ee
The expansion is in terms of conformal blocks of $SO(4,2)$ \cite{Dolan:2000ut},
\be \label{conformalblock}\notag
G^{(\ell)}_{\Delta} = \frac{1}{z-\zb}\Bigl((-\frac{1}{2} z)^\ell z f_{\Delta + \ell} (z) f_{\Delta - \ell -2} (\bar z) -   (z\leftrightarrow \zb) \Bigr)\, ,
\ee
with $f_\beta (z) ={}_2F_1 ( \beta/2, \beta/2 ; \beta ;z)$. The coefficients of the expansion are the squares of three-point coefficients, and so are required to be non-negative by unitarity. The function $\cG^{\rm long}$ receives contributions only from long multiplets of dimension $\Delta$ and even spin $\ell$, with $\Delta\geq\ell+2$ for unitarity \footnote{Each such multiplet contains a unique quasiprimary operator in the ${\bf 105}$ representation, with scaling dimension $\Delta +4$.}.

\begin{figure*}[t!]
\includegraphics[width=164pt]{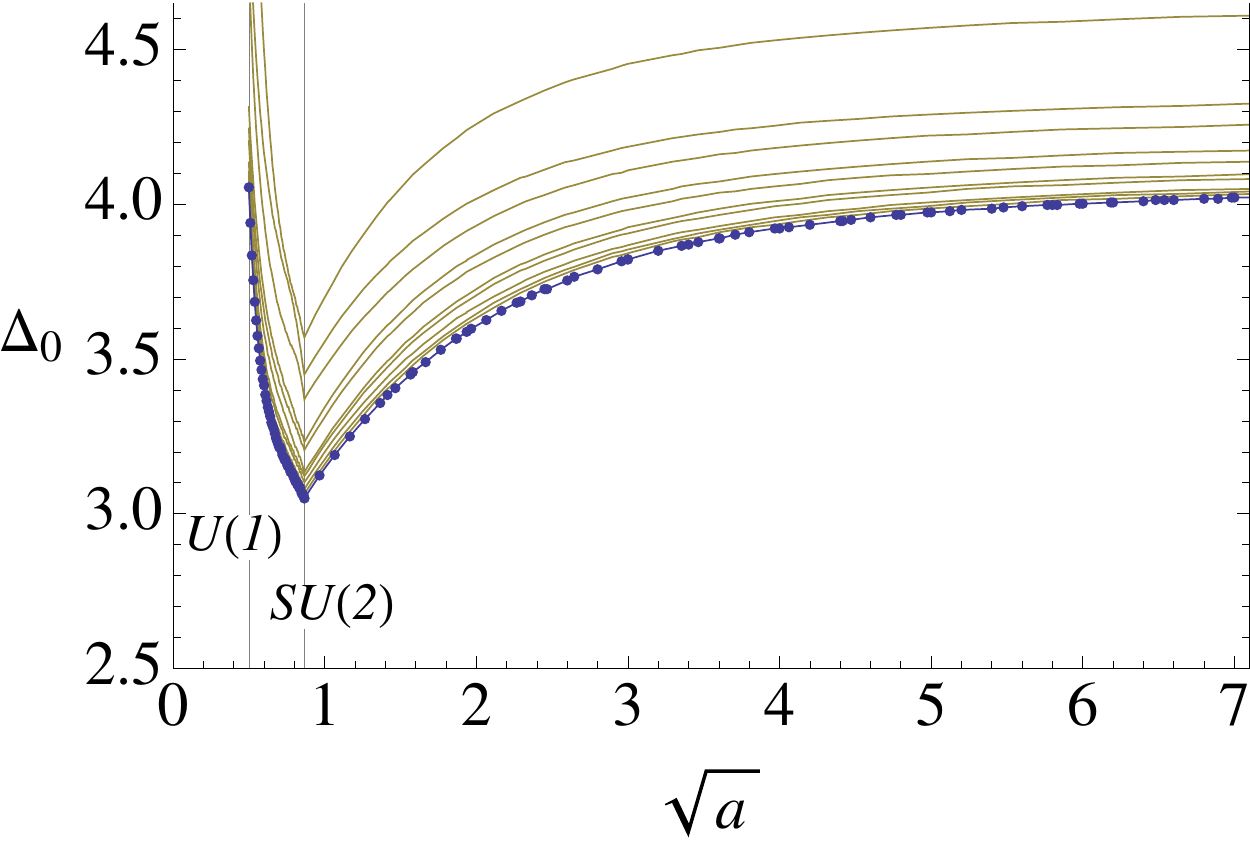}~~~
\includegraphics[width=164pt]{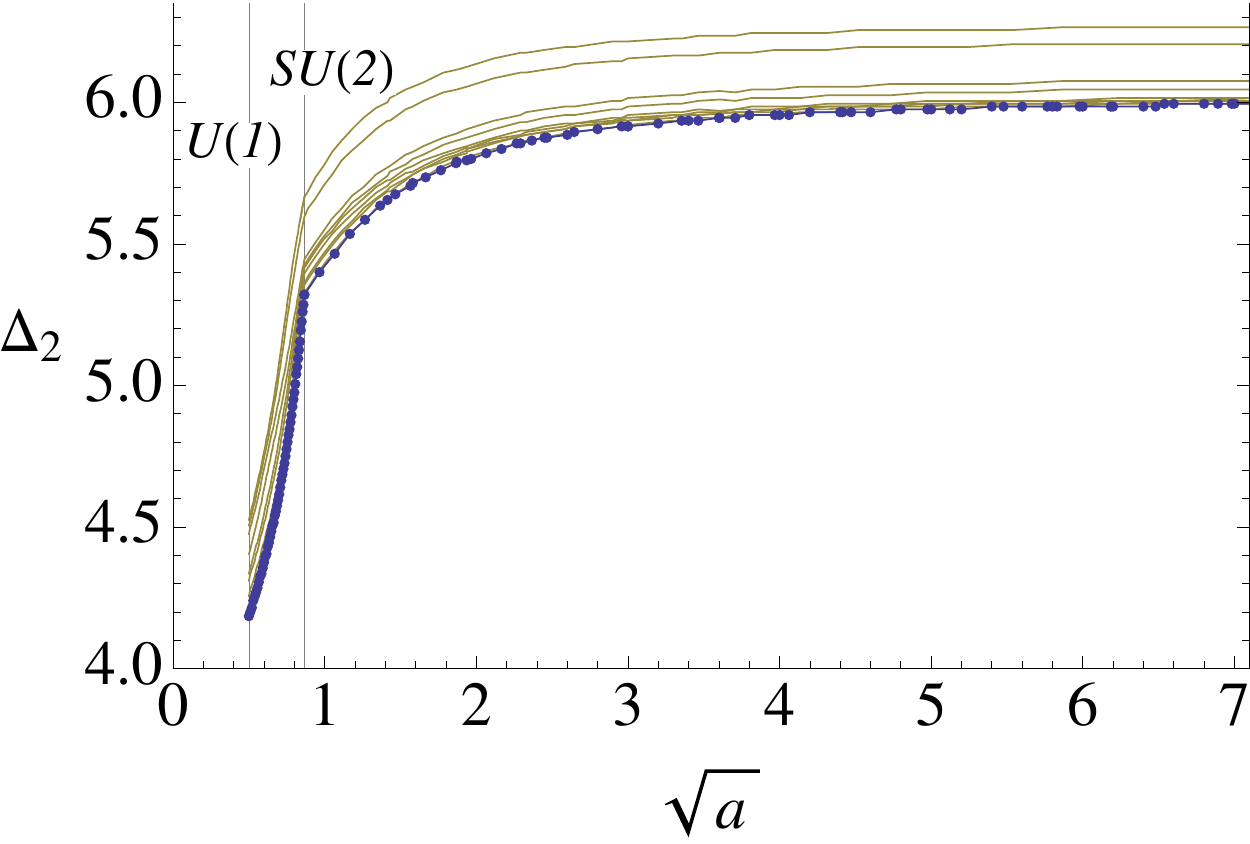}~~~
\includegraphics[width=164pt]{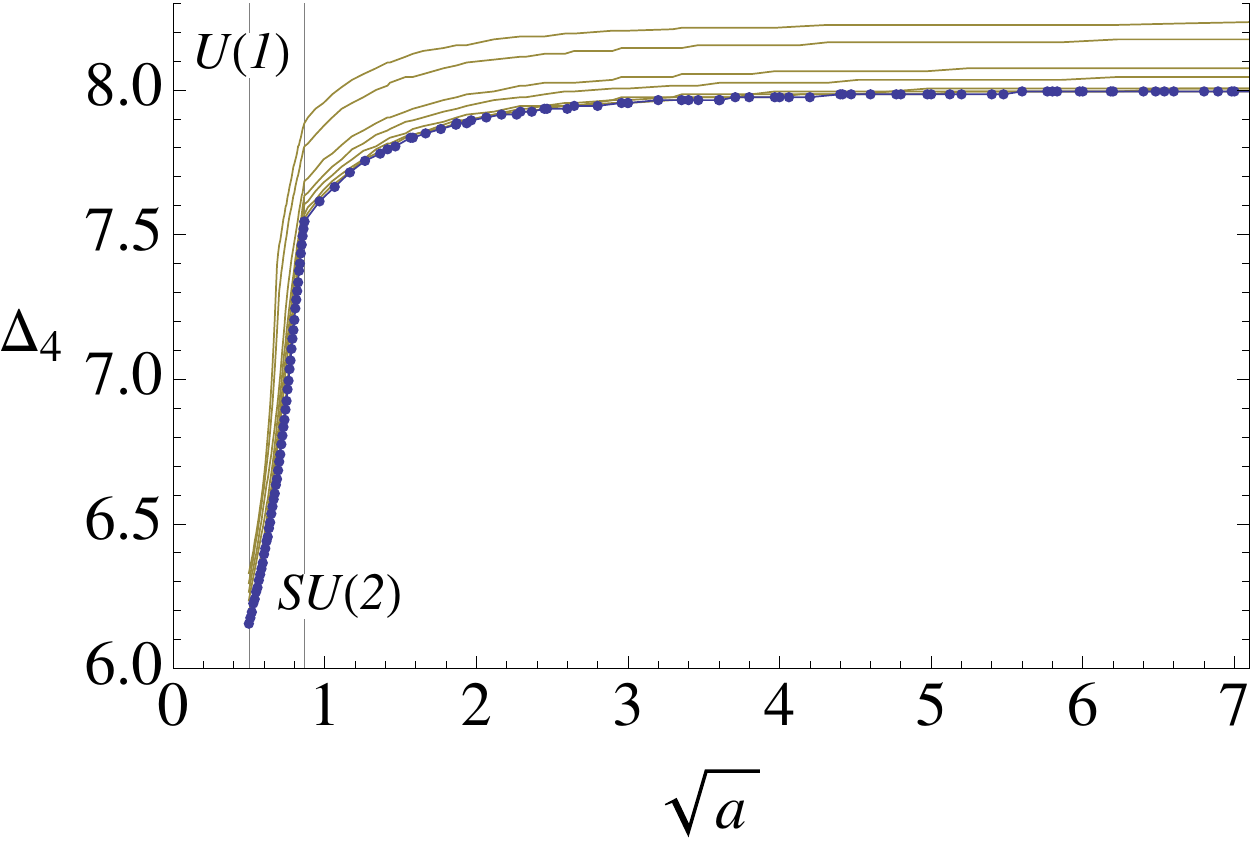}
\caption{Bounds for the scaling dimension of the leading twist unprotected operator of spin $\ell=0, 2, 4$. The bounds are displayed as a function of the (square root of the)  central charge $a$. The best bound is shown in blue, corresponding to $\Lambda=17$, while the lighter lines represent bounds for lower values of $\Lambda$.}
\label{fig:plots} 
\end{figure*}

The coefficients in the expansion of $\cG^{\rm short}$ are essentially determined by the protected parts of the amplitude. There is a potential ambiguity due to the possibility that operators in short representations may combine into a long representation at the unitarity bound.  However, there is always a canonical choice for the representations contributing to $\cG^{\rm short}$ such that unitarity amounts to non-negativity of the undetermined coefficients in $\cG^{\rm long}$. This choice is the same for all $a>3/4$, but jumps at $a=3/4$; for $a<3/4$ one must include a representation that contains higher spin conserved currents. This transition has a physical interpretation in $\cN=4$ SYM, where for $G=SU(2)$, a certain quarter-BPS operator is absent due to trace relations.

The crossing symmetry relation can now be recast as a sum rule for the unknown, non-negative coefficients $a_{\Delta,\ell}$,
\be\label{sumrule}\notag
\sum_{\Delta,\ell}a_{\Delta,\ell}F_{\Delta,\ell}(u,v)= F^{\rm short}(u,v;a)~,
\ee
where we have defined 
\be
F_{\Delta,\ell}(u,v)=v^2u^{\frac{\Delta-\ell}{2}}G^{(\ell)}_{\Delta+4}(u,v)-u^2v^{\frac{\Delta-\ell}{2}}G^{(\ell)}_{\Delta+4}(v,u)~,\nonumber
\ee
and $F^{\rm short}(u,v;a)$ is a known family of functions parametrized by the central charge $a$.

\vspace{10pt}
{\bf Numerical Constraints.}
Following \cite{Rattazzi:2008pe}, we now use the sum rule to place constraints on the spectrum of any $\cN=4$ SCFT. A trial spectrum can be ruled out if it is possible to find a real linear functional $\phi$ such that
\begin{enumerate}
\item[]\quad $\phi\cdot F_{\Delta,\ell}(u,v)\geq0$ \quad when\quad $a_{\Delta,\ell}\neq0$~,
\item[]\quad $\phi\cdot F^{\rm short}(u,v)<0$~.
\end{enumerate}
Entire families of trial spectra can be ruled out in this way by enforcing the same positivity conditions for continuous ranges of $\Delta$. We consider spectral ans\"atze that are parameterized by a lower bound $\Delta_\ell$ on the dimension of operators of spin $\ell$. (We focus here on $\ell=0,2,4$, but the strategy is easily generalized.) For a given choice of the $\Delta_\ell$, all such spectra will be ruled out if there exists a functional satisfying
\be\notag
\phi\cdot F_{\Delta,\ell}(u,v)\geq0 \quad \rm{for} \quad \Delta\geq\Delta_{\ell}~.
\ee
The goal is to map out the contour in $\Delta_\ell$ space that separates the spectra that can be ruled out with such a linear functional from those that cannot.

This problem is rendered tractable by choosing a finite-dimensional subspace of the infinite-dimensional space of linear functionals
consisting of operators of the form
\be\notag
\phi\cdot g(z,\bar z) := \sum_{m,n\leq\Lambda}    \alpha_{m,n} \, \restr{\del_z^m\del_{\bar z}^n g(z,\bar z)}{z=\bar z=1/2}~,
\ee
for an integer cutoff $\Lambda$ on the maximum number of derivatives. Furthermore, the space of conformal blocks -- normally a continuous family of functions labeled by $\Delta$ for each spin -- is discretized and truncated at a high but finite spin ($\ell \leq 20$ in the plots shown here) \footnote{The results reported here are largely independent of these approximations -- the bounds were found to not change significantly when the discretization was refined and the spin cutoff was increased.}. The task of finding a functional in this subspace that is positive for all of these conformal blocks then reduces to a linear programming problem and can be performed by a computer. We have used the \texttt{IBM ILOG CPLEX} optimizer interfaced with \texttt{Mathematica}.

The results are displayed in Figs. 1-3. The constraint surfaces for the triple $\{\Delta_0,\Delta_2,\Delta_4\}$ are displayed in \figr{fig:cubes} for several values of the central charge and with $\Lambda=17$. The results for all other values of the central charge that we have checked are qualitatively similar \footnote{\label{supp}A more extensive subset of our results can be found in \texttt{N4bootstrap.nb} in the supplemental materials for this submission. We have also included \texttt{bootstrap.mov}, which illustrates the convergence of the $a=3/4$ exclusion plot as the dimension of the search space is raised from $6$ to $72$.}. The area outside of a cube-shaped region is excluded.

Better resolution is obtained for a lower dimensional search space, and \figr{fig:plots} shows the bounds obtained by constraining the $\Delta_0$, $\Delta_2$, and $\Delta_4$ separately. The approximately cubic shape of the exclusion plots in \figr{fig:cubes} shows that the result for the bound of a given spin does not depend much on the choices of gaps $\Delta_\ell$ for the other spins.

\begin{figure*}[t!]
\includegraphics[width=166pt]{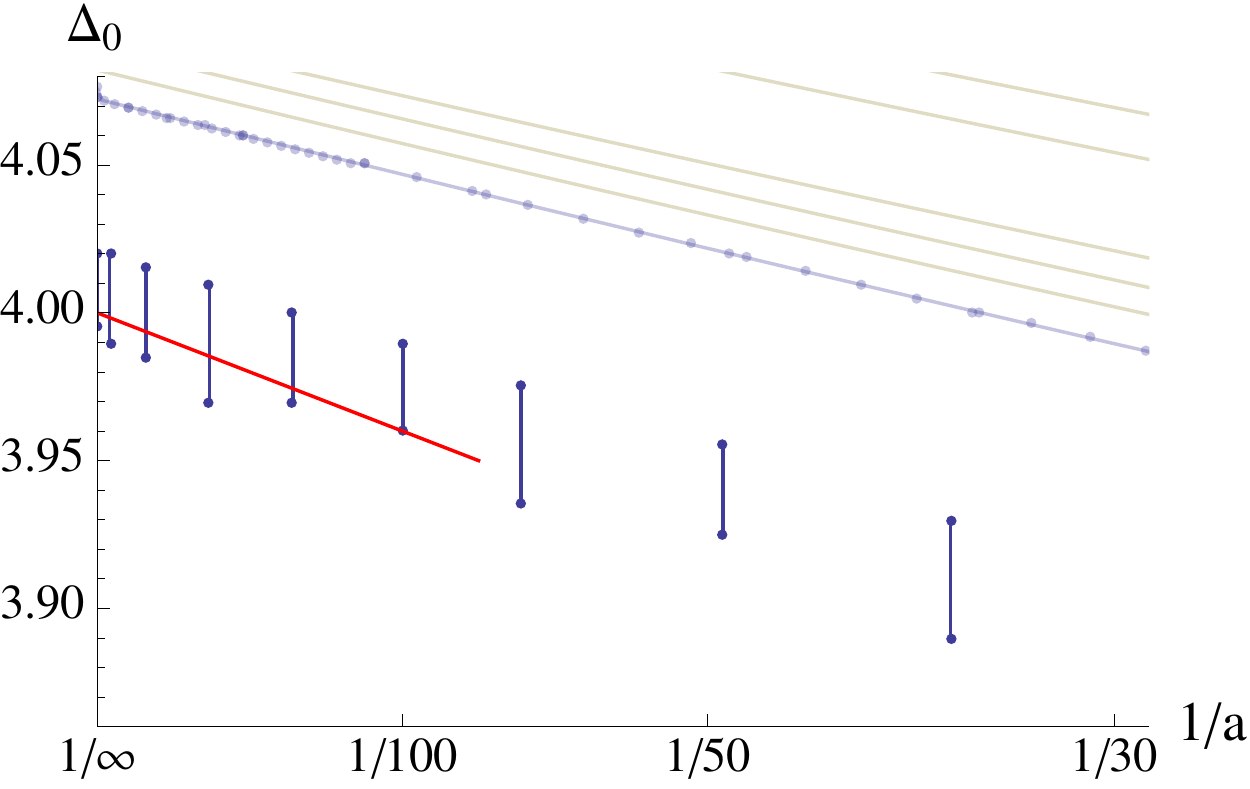}~~
\includegraphics[width=166pt]{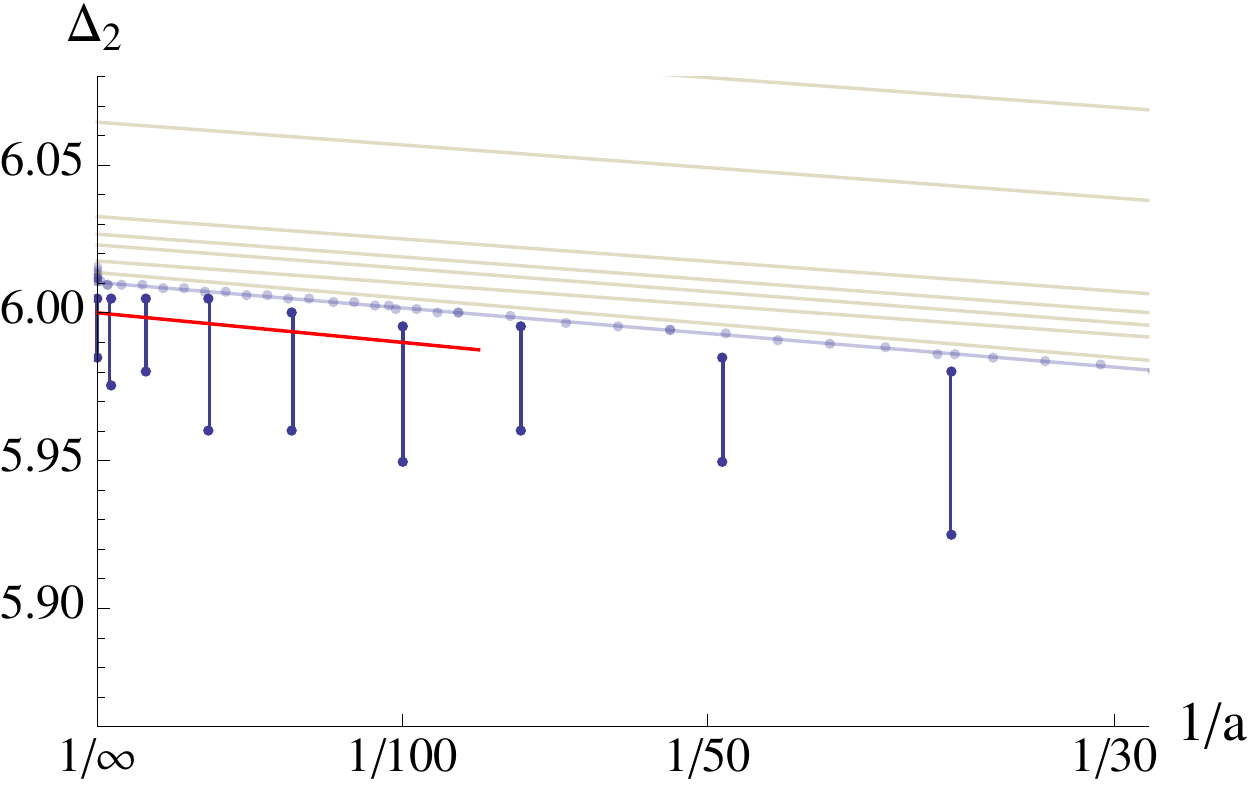}~~
\includegraphics[width=166pt]{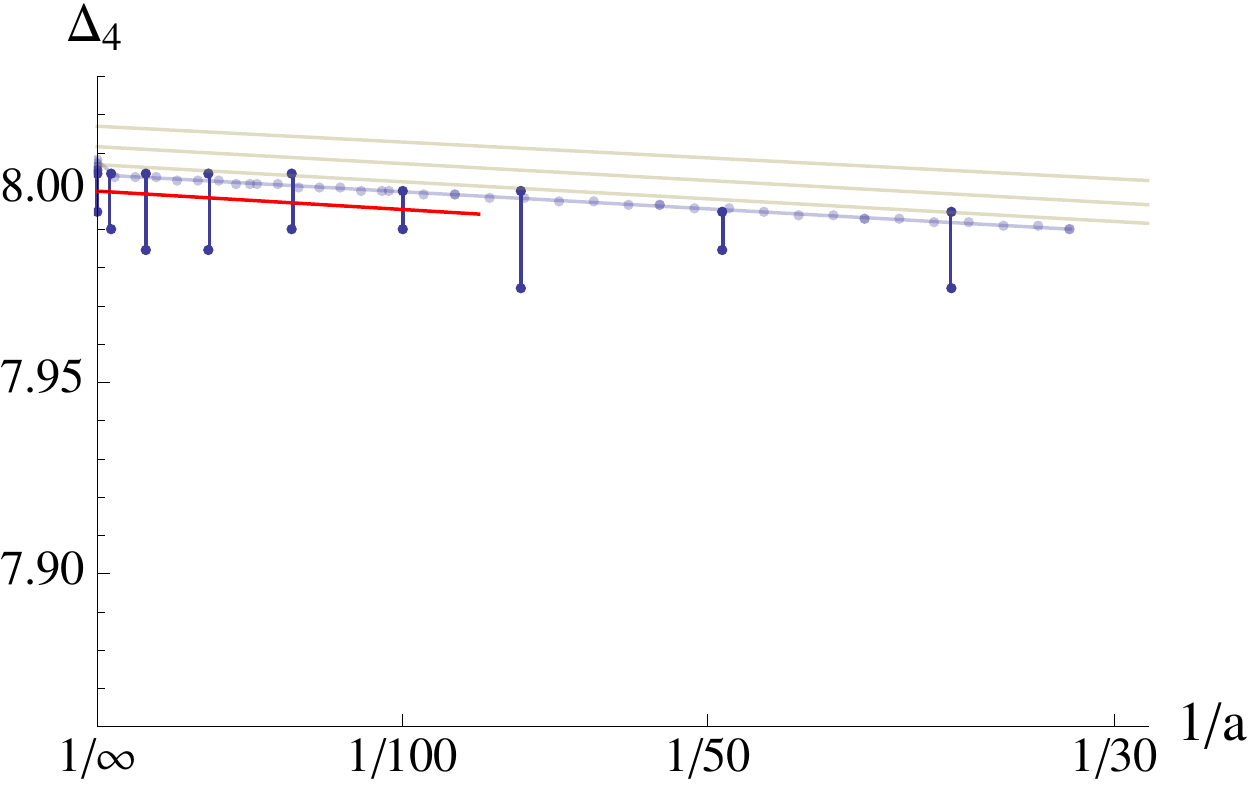}
\caption{Estimates for twist gap $\Delta_\ell$ for $\ell=0, 2, 4$ that characterize the corners of the exclusion cubes at large central charge. Uncertainty is due to the smoothing of the cube. The improvement relative to the single spin bounds of \figr{fig:plots} is apparent. Superimposed in red are the results for planar $\cN=4$ SYM in the limit of infinite 't Hooft coupling; $\Delta_0\approx4-\frac{4}{a}$, $\Delta_2\approx6-\frac 1a$, and $\Delta_4\approx8-\frac{12}{25a}$.}
\label{fig:cornerfit} 
\end{figure*}

Finally, \figr{fig:cornerfit} shows our best estimates for the values of the $\Delta_\ell$ at the corner of the cubic exclusions for large values of the central charge. We have superimposed the values of these dimensions to order $O(a\inv)$ for type IIB supergravity on $\AdS_5 \times S^5$.

\vspace{10pt}
{\bf Discussion.}
Even without additional interpretation, the results reported here provide true, nonperturbative bounds for the dimensions of unprotected operators in $\cN=4$ SYM for every gauge group. From the first graph of \figr{fig:plots} we can read off the bound for the dimension of the leading twist $SU(4)_R$ singlet scalar operator. Recall that in perturbative $\cN=4$ SYM with $G=SU(N)$, this operator is identified with the Konishi operator $\Tr X^iX^i$, with dimension $\Delta=2+\gamma(g_{\rm YM})$ (see, \eg, \cite{Bianchi:2001cm,*Eden:2012fe}). By contrast, in the planar limit at large 't Hooft coupling, the Konishi operator acquires a large anomalous dimension \cite{Gubser:1998bc,*Gromov:2009zb} and the leading twist operator is the double-trace operator $\cO^I_{\bf 20^\prime}\cO^I_{\bf 20^\prime}$, with dimension $\Delta \approx 4 - 16/N^2$ \cite{DHoker:1999jp,*Arutyunov:2000ku}. Our results provide a nonperturbative upper bound for any $N$. For example, the point corresponding to $a=3/4$ is marked in the plots of \figr{fig:plots} and gives bounds for $SU(2)$ SYM: $\Delta \leq 3.050$ for all values of the complexified gauge coupling. Similar results can be read off for any gauge groups and for spins zero, two, and four. These bounds can be made slightly stronger for a given spin in exchange for the assumption of a gap for the other spins. 

\begin{figure}[b!]
\includegraphics[width=185pt]{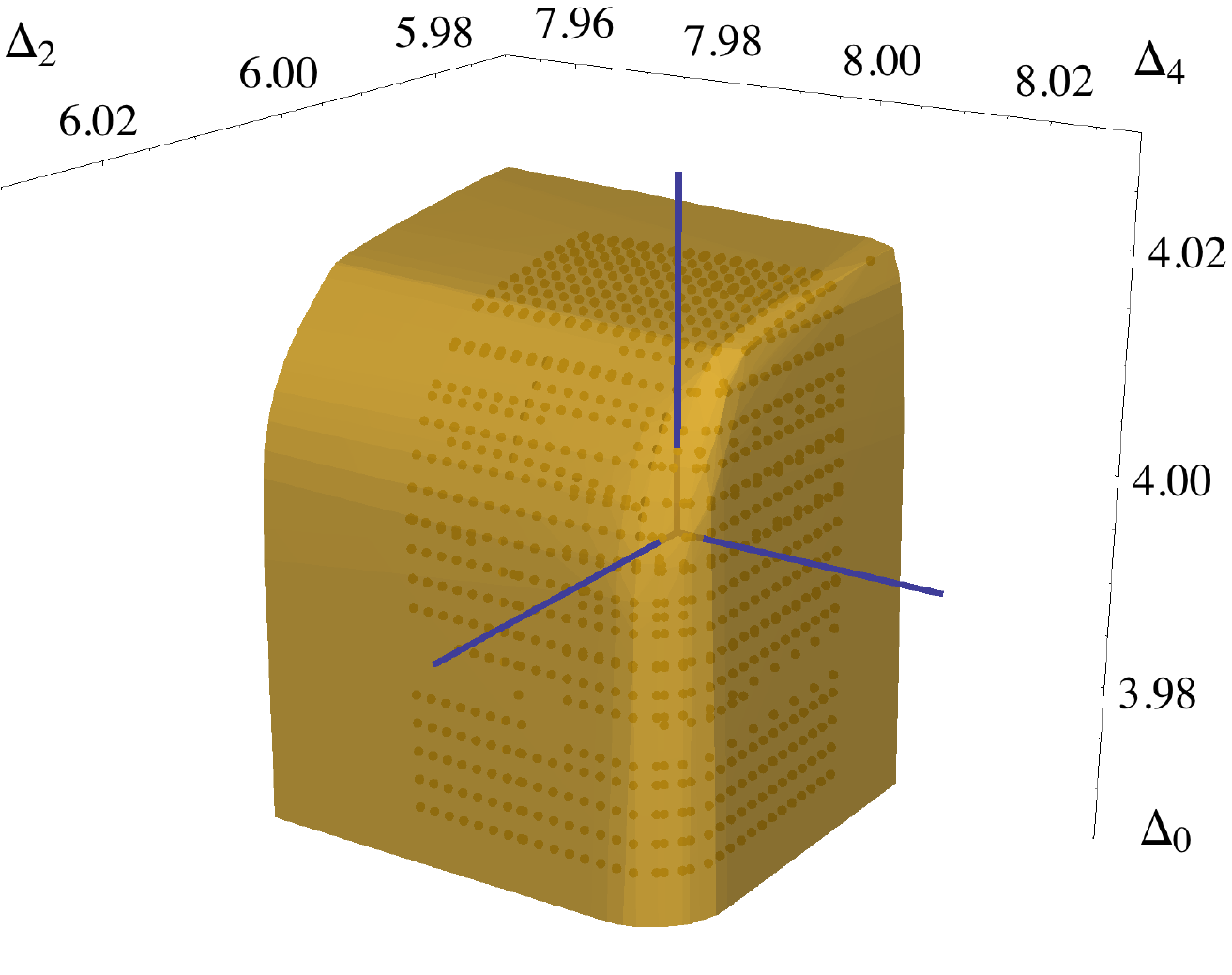}
\caption{A neighborhood of the corner of the exclusion boundary for infinite central charge. The origin of the superimposed axes sits at $\Delta_\ell = \ell+4$.}
\label{fig:cornerzoom} 
\end{figure}

There are, however, two features of our results that are suggestive of a more specific interpretation. The first is the behavior of the bounds for large values of the central charge (see Figs.\;\ref{fig:cornerfit},\;\ref{fig:cornerzoom}). For infinite central charge, the independent bounds for leading twist operators are
\be\notag
(\Delta_0,\Delta_2,\Delta_4)\alt(4.073, 6.011, 8.005)~.
\ee
These are very close to $\Delta_\ell=\ell+4$ -- the result that follows from large $N$ factorization for $\cN=4$ SYM, where the leading twist unprotected primaries are double-trace operators of the schematic form $\cO^{I}_{\bf 20^\prime} \partial^\ell \cO^{I}_{\bf 20^\prime}$. More significantly, the leading $1/a$ corrections to the bounds are compatible with the large 't Hooft coupling limit of planar SYM (as computed holographically in supergravity \cite{DHoker:1999jp,*Arutyunov:2000ku,Dolan:2001tt}). Additionally, the bounds at $a=1/4$ appear to converge to the correct values for $U(1)$ SYM \footnote{In the abelian theory, the contribution of the Konishi operator is included in $\cG^{\rm short}$, so the leading ``unprotected'' scalar is of dimension four. On the other hand, for $\ell>0$, the leading twist operators sit at the unitarity bound.}. \emph{A priori}, the true spectrum  of ${\cal N}=4$ SYM could have been impossible to access by  bootstrap methods, but we find instead evidence that our bounds are saturated by the physical theory both at large and small values of the central charge. We conjecture that this persists at all intermediate values.

The second hint comes from the most conspicuous feature of the plots in \figr{fig:cubes}; the boundary of the excluded region is approximately that of a cube. This is actually to be expected if the bounds owe their existence to an isolated solution to crossing symmetry with a spectrum of leading twist operators corresponding to the vertex of the cube \footnote{More precisely, the cubic shape  is expected in a finite neighborhood of the corner, and the functionals obtained for choices of gaps close to the corner values should all describe the same isolated solution to crossing symmetry.}. In the case at hand this requires further explanation, since four-point functions in $\cN=4$ SYM are not isolated, but rather vary continuously with the complexified gauge coupling. In order for a continuous family of four-point functions to generate a cube-shaped exclusion plot, it must be the case that for a single member of the family, the dimensions of the leading twist operators of all spins appearing in the OPE are maximized simultaneously.

As it happens, the conformal manifold of $\cN=4$ SYM has two special points that serve as stationary points for the dimensions of all operators in the theory. These are the $S$ and $S\cdot T$ invariant theories at $\tau=\theta/2\pi+4\pi i/g_{\rm YM}^2=\exp(\pi i/2)$ and $\tau=\exp(\pi i/3)$. This property follows from invariance under a finite subgroup of $SL(2,\mathbb Z)$. Since there is no clear reason for coincident extremization of anomalous dimensions at any other value of the coupling, the shape of the exclusion thresholds, along with the fact that the result matches known physics for extreme values of the central charge, are strong indications that the corners of the cube shaped boundaries in \figr{fig:cubes} can be used to approximate the spectrum at one of the self-dual points.

This is an extraordinary possibility, since there are no known tools for analyzing the self-dual points of $\cN=4$ SYM -- they are in fact the least accessible points to any existing methods. We are hopeful that by using additional data as input parameters, \eg, the dimension and OPE coefficient of an unprotected operator, it will be possible in future work to examine the entire conformal manifold of the theory.

\vspace{10pt}
\begin{acknowledgments}
We are grateful to the participants of the \emph{Back to the Bootstrap} workshops for many useful discussions and suggestions, and in particular to Sheer El-Showk for generously sharing his \texttt{CPLEX} interface. The work of C.B. is supported in part by DOE grant DE-FG02-92ER-40697. The work of B.v.R. and L.R. is partially supported by the NSF under Grants PHY-0969919 and PHY-0969739.
\end{acknowledgments}
\bibliography{bootstrap}

\begin{thebibliography}{31}%
\makeatletter
\providecommand \@ifxundefined [1]{%
 \@ifx{#1\undefined}
}%
\providecommand \@ifnum [1]{%
 \ifnum #1\expandafter \@firstoftwo
 \else \expandafter \@secondoftwo
 \fi
}%
\providecommand \@ifx [1]{%
 \ifx #1\expandafter \@firstoftwo
 \else \expandafter \@secondoftwo
 \fi
}%
\providecommand \natexlab [1]{#1}%
\providecommand \enquote  [1]{``#1''}%
\providecommand \bibnamefont  [1]{#1}%
\providecommand \bibfnamefont [1]{#1}%
\providecommand \citenamefont [1]{#1}%
\providecommand \href@noop [0]{\@secondoftwo}%
\providecommand \href [0]{\begingroup \@sanitize@url \@href}%
\providecommand \@href[1]{\@@startlink{#1}\@@href}%
\providecommand \@@href[1]{\endgroup#1\@@endlink}%
\providecommand \@sanitize@url [0]{\catcode `\\12\catcode `\$12\catcode
  `\&12\catcode `\#12\catcode `\^12\catcode `\_12\catcode `\%12\relax}%
\providecommand \@@startlink[1]{}%
\providecommand \@@endlink[0]{}%
\providecommand \url  [0]{\begingroup\@sanitize@url \@url }%
\providecommand \@url [1]{\endgroup\@href {#1}{\urlprefix }}%
\providecommand \urlprefix  [0]{URL }%
\providecommand \Eprint [0]{\href }%
\providecommand \doibase [0]{http://dx.doi.org/}%
\providecommand \selectlanguage [0]{\@gobble}%
\providecommand \bibinfo  [0]{\@secondoftwo}%
\providecommand \bibfield  [0]{\@secondoftwo}%
\providecommand \translation [1]{[#1]}%
\providecommand \BibitemOpen [0]{}%
\providecommand \bibitemStop [0]{}%
\providecommand \bibitemNoStop [0]{.\EOS\space}%
\providecommand \EOS [0]{\spacefactor3000\relax}%
\providecommand \BibitemShut  [1]{\csname bibitem#1\endcsname}%
\let\auto@bib@innerbib\@empty
\bibitem [{\citenamefont {Rattazzi}\ \emph {et~al.}(2008)\citenamefont
  {Rattazzi}, \citenamefont {Rychkov}, \citenamefont {Tonni},\ and\
  \citenamefont {Vichi}}]{Rattazzi:2008pe}%
  \BibitemOpen
  \bibfield  {author} {\bibinfo {author} {\bibfnamefont {R.}~\bibnamefont
  {Rattazzi}}, \bibinfo {author} {\bibfnamefont {V.~S.}\ \bibnamefont
  {Rychkov}}, \bibinfo {author} {\bibfnamefont {E.}~\bibnamefont {Tonni}}, \
  and\ \bibinfo {author} {\bibfnamefont {A.}~\bibnamefont {Vichi}},\ }\href
  {\doibase 10.1088/1126-6708/2008/12/031} {\bibfield  {journal} {\bibinfo
  {journal} {JHEP}\ }\textbf {\bibinfo {volume} {0812}},\ \bibinfo {pages}
  {031} (\bibinfo {year} {2008})},\ \Eprint {http://arxiv.org/abs/0807.0004}
  {arXiv:0807.0004 [hep-th]} \BibitemShut {NoStop}%
\bibitem [{\citenamefont {Rychkov}\ and\ \citenamefont
  {Vichi}(2009)}]{Rychkov:2009ij}%
  \BibitemOpen
  \bibfield  {author} {\bibinfo {author} {\bibfnamefont {V.~S.}\ \bibnamefont
  {Rychkov}}\ and\ \bibinfo {author} {\bibfnamefont {A.}~\bibnamefont
  {Vichi}},\ }\href {\doibase 10.1103/PhysRevD.80.045006} {\bibfield  {journal}
  {\bibinfo  {journal} {Phys.Rev.}\ }\textbf {\bibinfo {volume} {D80}},\
  \bibinfo {pages} {045006} (\bibinfo {year} {2009})},\ \Eprint
  {http://arxiv.org/abs/0905.2211} {arXiv:0905.2211 [hep-th]} \BibitemShut
  {NoStop}%
\bibitem [{\citenamefont {El-Showk}\ \emph {et~al.}(2012)\citenamefont
  {El-Showk}, \citenamefont {Paulos}, \citenamefont {Poland}, \citenamefont
  {Rychkov}, \citenamefont {Simmons-Duffin} \emph {et~al.}}]{ElShowk:2012ht}%
  \BibitemOpen
  \bibfield  {author} {\bibinfo {author} {\bibfnamefont {S.}~\bibnamefont
  {El-Showk}}, \bibinfo {author} {\bibfnamefont {M.~F.}\ \bibnamefont
  {Paulos}}, \bibinfo {author} {\bibfnamefont {D.}~\bibnamefont {Poland}},
  \bibinfo {author} {\bibfnamefont {S.}~\bibnamefont {Rychkov}}, \bibinfo
  {author} {\bibfnamefont {D.}~\bibnamefont {Simmons-Duffin}},  \emph
  {et~al.},\ }\href {\doibase 10.1103/PhysRevD.86.025022} {\bibfield  {journal}
  {\bibinfo  {journal} {Phys.Rev.}\ }\textbf {\bibinfo {volume} {D86}},\
  \bibinfo {pages} {025022} (\bibinfo {year} {2012})},\ \Eprint
  {http://arxiv.org/abs/1203.6064} {arXiv:1203.6064 [hep-th]} \BibitemShut
  {NoStop}%
\bibitem [{\citenamefont {Poland}\ and\ \citenamefont
  {Simmons-Duffin}(2011)}]{Poland:2010wg}%
  \BibitemOpen
  \bibfield  {author} {\bibinfo {author} {\bibfnamefont {D.}~\bibnamefont
  {Poland}}\ and\ \bibinfo {author} {\bibfnamefont {D.}~\bibnamefont
  {Simmons-Duffin}},\ }\href {\doibase 10.1007/JHEP05(2011)017} {\bibfield
  {journal} {\bibinfo  {journal} {JHEP}\ }\textbf {\bibinfo {volume} {1105}},\
  \bibinfo {pages} {017} (\bibinfo {year} {2011})},\ \Eprint
  {http://arxiv.org/abs/1009.2087} {arXiv:1009.2087 [hep-th]} \BibitemShut
  {NoStop}%
\bibitem [{\citenamefont {El-Showk}\ and\ \citenamefont
  {Paulos}(2012)}]{ElShowk:2012hu}%
  \BibitemOpen
  \bibfield  {author} {\bibinfo {author} {\bibfnamefont {S.}~\bibnamefont
  {El-Showk}}\ and\ \bibinfo {author} {\bibfnamefont {M.~F.}\ \bibnamefont
  {Paulos}},\ }\href@noop {} {\  (\bibinfo {year} {2012})},\ \Eprint
  {http://arxiv.org/abs/1211.2810} {arXiv:1211.2810 [hep-th]} \BibitemShut
  {NoStop}%
\bibitem [{Note1()}]{Note1}%
  \BibitemOpen
  \bibinfo {note} {A more complete account of our methods, along with
  additional results, will appear in \cite {BRR:long}.}\BibitemShut {Stop}%
\bibitem [{\citenamefont {Freedman}\ \emph {et~al.}(1992)\citenamefont
  {Freedman}, \citenamefont {Johnson},\ and\ \citenamefont
  {Latorre}}]{Freedman:1991tk}%
  \BibitemOpen
  \bibfield  {author} {\bibinfo {author} {\bibfnamefont {D.~Z.}\ \bibnamefont
  {Freedman}}, \bibinfo {author} {\bibfnamefont {K.}~\bibnamefont {Johnson}}, \
  and\ \bibinfo {author} {\bibfnamefont {J.~I.}\ \bibnamefont {Latorre}},\
  }\href {\doibase 10.1016/0550-3213(92)90240-C} {\bibfield  {journal}
  {\bibinfo  {journal} {Nucl.Phys.}\ }\textbf {\bibinfo {volume} {B371}},\
  \bibinfo {pages} {353} (\bibinfo {year} {1992})}\BibitemShut {NoStop}%
\bibitem [{\citenamefont {Usyukina}\ and\ \citenamefont
  {Davydychev}(1993{\natexlab{a}})}]{Usyukina:1992jd}%
  \BibitemOpen
  \bibfield  {author} {\bibinfo {author} {\bibfnamefont {N.}~\bibnamefont
  {Usyukina}}\ and\ \bibinfo {author} {\bibfnamefont {A.~I.}\ \bibnamefont
  {Davydychev}},\ }\href {\doibase 10.1016/0370-2693(93)91834-A} {\bibfield
  {journal} {\bibinfo  {journal} {Phys.Lett.}\ }\textbf {\bibinfo {volume}
  {B298}},\ \bibinfo {pages} {363} (\bibinfo {year}
  {1993}{\natexlab{a}})}\BibitemShut {NoStop}%
\bibitem [{\citenamefont {Usyukina}\ and\ \citenamefont
  {Davydychev}(1993{\natexlab{b}})}]{Usyukina:1993ch}%
  \BibitemOpen
  \bibfield  {author} {\bibinfo {author} {\bibfnamefont {N.}~\bibnamefont
  {Usyukina}}\ and\ \bibinfo {author} {\bibfnamefont {A.~I.}\ \bibnamefont
  {Davydychev}},\ }\href {\doibase 10.1016/0370-2693(93)91118-7} {\bibfield
  {journal} {\bibinfo  {journal} {Phys.Lett.}\ }\textbf {\bibinfo {volume}
  {B305}},\ \bibinfo {pages} {136} (\bibinfo {year}
  {1993}{\natexlab{b}})}\BibitemShut {NoStop}%
\bibitem [{\citenamefont {Dolan}\ and\ \citenamefont
  {Osborn}(2001)}]{Dolan:2000ut}%
  \BibitemOpen
  \bibfield  {author} {\bibinfo {author} {\bibfnamefont {F.}~\bibnamefont
  {Dolan}}\ and\ \bibinfo {author} {\bibfnamefont {H.}~\bibnamefont {Osborn}},\
  }\href {\doibase 10.1016/S0550-3213(01)00013-X} {\bibfield  {journal}
  {\bibinfo  {journal} {Nucl.Phys.}\ }\textbf {\bibinfo {volume} {B599}},\
  \bibinfo {pages} {459} (\bibinfo {year} {2001})},\ \Eprint
  {http://arxiv.org/abs/hep-th/0011040} {arXiv:hep-th/0011040 [hep-th]}
  \BibitemShut {NoStop}%
\bibitem [{\citenamefont {Eden}\ \emph {et~al.}(2001)\citenamefont {Eden},
  \citenamefont {Petkou}, \citenamefont {Schubert},\ and\ \citenamefont
  {Sokatchev}}]{Eden:2000bk}%
  \BibitemOpen
  \bibfield  {author} {\bibinfo {author} {\bibfnamefont {B.}~\bibnamefont
  {Eden}}, \bibinfo {author} {\bibfnamefont {A.~C.}\ \bibnamefont {Petkou}},
  \bibinfo {author} {\bibfnamefont {C.}~\bibnamefont {Schubert}}, \ and\
  \bibinfo {author} {\bibfnamefont {E.}~\bibnamefont {Sokatchev}},\ }\href
  {\doibase 10.1016/S0550-3213(01)00151-1} {\bibfield  {journal} {\bibinfo
  {journal} {Nucl.Phys.}\ }\textbf {\bibinfo {volume} {B607}},\ \bibinfo
  {pages} {191} (\bibinfo {year} {2001})},\ \Eprint
  {http://arxiv.org/abs/hep-th/0009106} {arXiv:hep-th/0009106 [hep-th]}
  \BibitemShut {NoStop}%
\bibitem [{\citenamefont {Eden}\ and\ \citenamefont
  {Sokatchev}(2001)}]{Eden:2001ec}%
  \BibitemOpen
  \bibfield  {author} {\bibinfo {author} {\bibfnamefont {B.}~\bibnamefont
  {Eden}}\ and\ \bibinfo {author} {\bibfnamefont {E.}~\bibnamefont
  {Sokatchev}},\ }\href {\doibase 10.1016/S0550-3213(01)00492-8} {\bibfield
  {journal} {\bibinfo  {journal} {Nucl.Phys.}\ }\textbf {\bibinfo {volume}
  {B618}},\ \bibinfo {pages} {259} (\bibinfo {year} {2001})},\ \Eprint
  {http://arxiv.org/abs/hep-th/0106249} {arXiv:hep-th/0106249 [hep-th]}
  \BibitemShut {NoStop}%
\bibitem [{\citenamefont {Dolan}\ and\ \citenamefont
  {Osborn}(2002)}]{Dolan:2001tt}%
  \BibitemOpen
  \bibfield  {author} {\bibinfo {author} {\bibfnamefont {F.}~\bibnamefont
  {Dolan}}\ and\ \bibinfo {author} {\bibfnamefont {H.}~\bibnamefont {Osborn}},\
  }\href {\doibase 10.1016/S0550-3213(02)00096-2} {\bibfield  {journal}
  {\bibinfo  {journal} {Nucl.Phys.}\ }\textbf {\bibinfo {volume} {B629}},\
  \bibinfo {pages} {3} (\bibinfo {year} {2002})},\ \Eprint
  {http://arxiv.org/abs/hep-th/0112251} {arXiv:hep-th/0112251 [hep-th]}
  \BibitemShut {NoStop}%
\bibitem [{\citenamefont {Heslop}\ and\ \citenamefont
  {Howe}(2003)}]{Heslop:2002hp}%
  \BibitemOpen
  \bibfield  {author} {\bibinfo {author} {\bibfnamefont {P.}~\bibnamefont
  {Heslop}}\ and\ \bibinfo {author} {\bibfnamefont {P.}~\bibnamefont {Howe}},\
  }\href@noop {} {\bibfield  {journal} {\bibinfo  {journal} {JHEP}\ }\textbf
  {\bibinfo {volume} {0301}},\ \bibinfo {pages} {043} (\bibinfo {year}
  {2003})},\ \Eprint {http://arxiv.org/abs/hep-th/0211252}
  {arXiv:hep-th/0211252 [hep-th]} \BibitemShut {NoStop}%
\bibitem [{\citenamefont {Dolan}\ \emph {et~al.}(2004)\citenamefont {Dolan},
  \citenamefont {Gallot},\ and\ \citenamefont {Sokatchev}}]{Dolan:2004mu}%
  \BibitemOpen
  \bibfield  {author} {\bibinfo {author} {\bibfnamefont {F.~A.}\ \bibnamefont
  {Dolan}}, \bibinfo {author} {\bibfnamefont {L.}~\bibnamefont {Gallot}}, \
  and\ \bibinfo {author} {\bibfnamefont {E.}~\bibnamefont {Sokatchev}},\ }\href
  {\doibase 10.1088/1126-6708/2004/09/056} {\bibfield  {journal} {\bibinfo
  {journal} {JHEP}\ }\textbf {\bibinfo {volume} {0409}},\ \bibinfo {pages}
  {056} (\bibinfo {year} {2004})},\ \Eprint
  {http://arxiv.org/abs/hep-th/0405180} {arXiv:hep-th/0405180 [hep-th]}
  \BibitemShut {NoStop}%
\bibitem [{\citenamefont {Nirschl}\ and\ \citenamefont
  {Osborn}(2005)}]{Nirschl:2004pa}%
  \BibitemOpen
  \bibfield  {author} {\bibinfo {author} {\bibfnamefont {M.}~\bibnamefont
  {Nirschl}}\ and\ \bibinfo {author} {\bibfnamefont {H.}~\bibnamefont
  {Osborn}},\ }\href {\doibase 10.1016/j.nuclphysb.2005.01.013} {\bibfield
  {journal} {\bibinfo  {journal} {Nucl.Phys.}\ }\textbf {\bibinfo {volume}
  {B711}},\ \bibinfo {pages} {409} (\bibinfo {year} {2005})},\ \Eprint
  {http://arxiv.org/abs/hep-th/0407060} {arXiv:hep-th/0407060 [hep-th]}
  \BibitemShut {NoStop}%
\bibitem [{Note2()}]{Note2}%
  \BibitemOpen
  \bibinfo {note} {The constraints of crossing symmetry on the spectrum of
  protected operators in ${\protect \cal N}=4$ and ${\protect \cal N}=2$ SCFTs
  is a beautiful subject that will be elaborated upon in \cite
  {miniboot}.}\BibitemShut {Stop}%
\bibitem [{\citenamefont {Arutyunov}\ \emph {et~al.}(2002)\citenamefont
  {Arutyunov}, \citenamefont {Eden}, \citenamefont {Petkou},\ and\
  \citenamefont {Sokatchev}}]{Arutyunov:2001mh}%
  \BibitemOpen
  \bibfield  {author} {\bibinfo {author} {\bibfnamefont {G.}~\bibnamefont
  {Arutyunov}}, \bibinfo {author} {\bibfnamefont {B.}~\bibnamefont {Eden}},
  \bibinfo {author} {\bibfnamefont {A.}~\bibnamefont {Petkou}}, \ and\ \bibinfo
  {author} {\bibfnamefont {E.}~\bibnamefont {Sokatchev}},\ }\href {\doibase
  10.1016/S0550-3213(01)00569-7} {\bibfield  {journal} {\bibinfo  {journal}
  {Nucl.Phys.}\ }\textbf {\bibinfo {volume} {B620}},\ \bibinfo {pages} {380}
  (\bibinfo {year} {2002})},\ \Eprint {http://arxiv.org/abs/hep-th/0103230}
  {arXiv:hep-th/0103230 [hep-th]} \BibitemShut {NoStop}%
\bibitem [{Note3()}]{Note3}%
  \BibitemOpen
  \bibinfo {note} {Each such multiplet contains a unique quasiprimary operator
  in the ${\protect \bf 105}$ representation, with scaling dimension $\Delta
  +4$.}\BibitemShut {Stop}%
\bibitem [{Note4()}]{Note4}%
  \BibitemOpen
  \bibinfo {note} {The results reported here are largely independent of these
  approximations -- the bounds were found to not change significantly when the
  discretization was refined and the spin cutoff was increased.}\BibitemShut
  {Stop}%
\bibitem [{Note5()}]{Note5}%
  \BibitemOpen
  \bibinfo {note} {\label {supp}A more extensive subset of our results can be
  found in \protect \texttt {N4bootstrap.nb} in the supplemental materials for
  this submission. We have also included \protect \texttt {bootstrap.mov},
  which illustrates the convergence of the $a=3/4$ exclusion plot as the
  dimension of the search space is raised from $6$ to $72$.}\BibitemShut
  {Stop}%
\bibitem [{\citenamefont {Bianchi}\ \emph {et~al.}(2001)\citenamefont
  {Bianchi}, \citenamefont {Kovacs}, \citenamefont {Rossi},\ and\ \citenamefont
  {Stanev}}]{Bianchi:2001cm}%
  \BibitemOpen
  \bibfield  {author} {\bibinfo {author} {\bibfnamefont {M.}~\bibnamefont
  {Bianchi}}, \bibinfo {author} {\bibfnamefont {S.}~\bibnamefont {Kovacs}},
  \bibinfo {author} {\bibfnamefont {G.}~\bibnamefont {Rossi}}, \ and\ \bibinfo
  {author} {\bibfnamefont {Y.~S.}\ \bibnamefont {Stanev}},\ }\href@noop {}
  {\bibfield  {journal} {\bibinfo  {journal} {JHEP}\ }\textbf {\bibinfo
  {volume} {0105}},\ \bibinfo {pages} {042} (\bibinfo {year} {2001})},\ \Eprint
  {http://arxiv.org/abs/hep-th/0104016} {arXiv:hep-th/0104016 [hep-th]}
  \BibitemShut {NoStop}%
\bibitem [{\citenamefont {Eden}\ \emph {et~al.}(2012)\citenamefont {Eden},
  \citenamefont {Heslop}, \citenamefont {Korchemsky}, \citenamefont {Smirnov},\
  and\ \citenamefont {Sokatchev}}]{Eden:2012fe}%
  \BibitemOpen
  \bibfield  {author} {\bibinfo {author} {\bibfnamefont {B.}~\bibnamefont
  {Eden}}, \bibinfo {author} {\bibfnamefont {P.}~\bibnamefont {Heslop}},
  \bibinfo {author} {\bibfnamefont {G.~P.}\ \bibnamefont {Korchemsky}},
  \bibinfo {author} {\bibfnamefont {V.~A.}\ \bibnamefont {Smirnov}}, \ and\
  \bibinfo {author} {\bibfnamefont {E.}~\bibnamefont {Sokatchev}},\ }\href
  {\doibase 10.1016/j.nuclphysb.2012.04.015} {\bibfield  {journal} {\bibinfo
  {journal} {Nucl.Phys.}\ }\textbf {\bibinfo {volume} {B862}},\ \bibinfo
  {pages} {123} (\bibinfo {year} {2012})},\ \Eprint
  {http://arxiv.org/abs/1202.5733} {arXiv:1202.5733 [hep-th]} \BibitemShut
  {NoStop}%
\bibitem [{\citenamefont {Gubser}\ \emph {et~al.}(1998)\citenamefont {Gubser},
  \citenamefont {Klebanov},\ and\ \citenamefont {Polyakov}}]{Gubser:1998bc}%
  \BibitemOpen
  \bibfield  {author} {\bibinfo {author} {\bibfnamefont {S.}~\bibnamefont
  {Gubser}}, \bibinfo {author} {\bibfnamefont {I.~R.}\ \bibnamefont
  {Klebanov}}, \ and\ \bibinfo {author} {\bibfnamefont {A.~M.}\ \bibnamefont
  {Polyakov}},\ }\href {\doibase 10.1016/S0370-2693(98)00377-3} {\bibfield
  {journal} {\bibinfo  {journal} {Phys.Lett.}\ }\textbf {\bibinfo {volume}
  {B428}},\ \bibinfo {pages} {105} (\bibinfo {year} {1998})},\ \Eprint
  {http://arxiv.org/abs/hep-th/9802109} {arXiv:hep-th/9802109 [hep-th]}
  \BibitemShut {NoStop}%
\bibitem [{\citenamefont {Gromov}\ \emph {et~al.}(2010)\citenamefont {Gromov},
  \citenamefont {Kazakov},\ and\ \citenamefont {Vieira}}]{Gromov:2009zb}%
  \BibitemOpen
  \bibfield  {author} {\bibinfo {author} {\bibfnamefont {N.}~\bibnamefont
  {Gromov}}, \bibinfo {author} {\bibfnamefont {V.}~\bibnamefont {Kazakov}}, \
  and\ \bibinfo {author} {\bibfnamefont {P.}~\bibnamefont {Vieira}},\ }\href
  {\doibase 10.1103/PhysRevLett.104.211601} {\bibfield  {journal} {\bibinfo
  {journal} {Phys.Rev.Lett.}\ }\textbf {\bibinfo {volume} {104}},\ \bibinfo
  {pages} {211601} (\bibinfo {year} {2010})},\ \Eprint
  {http://arxiv.org/abs/0906.4240} {arXiv:0906.4240 [hep-th]} \BibitemShut
  {NoStop}%
\bibitem [{\citenamefont {D'Hoker}\ \emph {et~al.}(2000)\citenamefont
  {D'Hoker}, \citenamefont {Mathur}, \citenamefont {Matusis},\ and\
  \citenamefont {Rastelli}}]{DHoker:1999jp}%
  \BibitemOpen
  \bibfield  {author} {\bibinfo {author} {\bibfnamefont {E.}~\bibnamefont
  {D'Hoker}}, \bibinfo {author} {\bibfnamefont {S.~D.}\ \bibnamefont {Mathur}},
  \bibinfo {author} {\bibfnamefont {A.}~\bibnamefont {Matusis}}, \ and\
  \bibinfo {author} {\bibfnamefont {L.}~\bibnamefont {Rastelli}},\ }\href
  {\doibase 10.1016/S0550-3213(00)00523-X} {\bibfield  {journal} {\bibinfo
  {journal} {Nucl.Phys.}\ }\textbf {\bibinfo {volume} {B589}},\ \bibinfo
  {pages} {38} (\bibinfo {year} {2000})},\ \Eprint
  {http://arxiv.org/abs/hep-th/9911222} {arXiv:hep-th/9911222 [hep-th]}
  \BibitemShut {NoStop}%
\bibitem [{\citenamefont {Arutyunov}\ \emph {et~al.}(2000)\citenamefont
  {Arutyunov}, \citenamefont {Frolov},\ and\ \citenamefont
  {Petkou}}]{Arutyunov:2000ku}%
  \BibitemOpen
  \bibfield  {author} {\bibinfo {author} {\bibfnamefont {G.}~\bibnamefont
  {Arutyunov}}, \bibinfo {author} {\bibfnamefont {S.}~\bibnamefont {Frolov}}, \
  and\ \bibinfo {author} {\bibfnamefont {A.~C.}\ \bibnamefont {Petkou}},\
  }\href {\doibase 10.1016/S0550-3213(00)00439-9} {\bibfield  {journal}
  {\bibinfo  {journal} {Nucl.Phys.}\ }\textbf {\bibinfo {volume} {B586}},\
  \bibinfo {pages} {547} (\bibinfo {year} {2000})},\ \Eprint
  {http://arxiv.org/abs/hep-th/0005182} {arXiv:hep-th/0005182 [hep-th]}
  \BibitemShut {NoStop}%
\bibitem [{Note6()}]{Note6}%
  \BibitemOpen
  \bibinfo {note} {In the abelian theory, the contribution of the Konishi
  operator is included in ${\protect \cal G}^{\protect \rm short}$, so the
  leading ``unprotected'' scalar is of dimension four. On the other hand, for
  $\ell >0$, the leading twist operators sit at the unitarity
  bound.}\BibitemShut {Stop}%
\bibitem [{Note7()}]{Note7}%
  \BibitemOpen
  \bibinfo {note} {More precisely, the cubic shape is expected in a finite
  neighborhood of the corner, and the functionals obtained for choices of gaps
  close to the corner values should all describe the same isolated solution to
  crossing symmetry.}\BibitemShut {Stop}%
\bibitem [{\citenamefont {Beem}\ \emph {et~al.}(pear)\citenamefont {Beem},
  \citenamefont {Rastelli},\ and\ \citenamefont {van Rees}}]{BRR:long}%
  \BibitemOpen
  \bibfield  {author} {\bibinfo {author} {\bibfnamefont {C.}~\bibnamefont
  {Beem}}, \bibinfo {author} {\bibfnamefont {L.}~\bibnamefont {Rastelli}}, \
  and\ \bibinfo {author} {\bibfnamefont {B.~C.}\ \bibnamefont {van Rees}},\
  }\href@noop {} {\  (\bibinfo {year} {\emph{to appear}})}\BibitemShut
  {NoStop}%
\bibitem [{\citenamefont {Beem}\ \emph {et~al.}(ress)\citenamefont {Beem},
  \citenamefont {Lemos}, \citenamefont {Liendo}, \citenamefont {Peelaers},
  \citenamefont {Rastelli},\ and\ \citenamefont {van Rees}}]{miniboot}%
  \BibitemOpen
  \bibfield  {author} {\bibinfo {author} {\bibfnamefont {C.}~\bibnamefont
  {Beem}}, \bibinfo {author} {\bibfnamefont {M.}~\bibnamefont {Lemos}},
  \bibinfo {author} {\bibfnamefont {P.}~\bibnamefont {Liendo}}, \bibinfo
  {author} {\bibfnamefont {W.}~\bibnamefont {Peelaers}}, \bibinfo {author}
  {\bibfnamefont {L.}~\bibnamefont {Rastelli}}, \ and\ \bibinfo {author}
  {\bibfnamefont {B.~C.}\ \bibnamefont {van Rees}},\ }\href@noop {} {\
  (\bibinfo {year} {\emph{work in progress}})}\BibitemShut {NoStop}%
\end{thebibliography}%
\end{document}